\documentclass[11pt,a4paper]{article}
\pdfoutput=1
\usepackage{jcappub}
\usepackage[T1]{fontenc} 

\usepackage{graphicx,natbib}
\usepackage{dcolumn}
\usepackage{bm}
\usepackage{hyperref}
\usepackage{color}
\usepackage{soul}

\RequirePackage[english]{babel}
\usepackage{bm}      
\usepackage{amssymb}   
\usepackage{amsmath}   
\usepackage{graphicx}	
\usepackage{nicefrac}

\begin{document}
\title{Hybrid stars in the light of the merging event GW170817}

\author[1]{Alessandro Parisi,\href{https://orcid.org/0000-0003-0251-8914}{\includegraphics[scale=0.1]{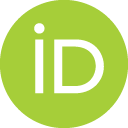}}}
\author[2]{C. V\'{a}squez Flores,}
\author[3]{C. Henrique Lenzi,\href{https://orcid.org/0000-0001-5887-338X}{\includegraphics[scale=0.1]{ORCIDiD_icon128x128.png}}}
\author[1]{Chian-Shu Chen}
\author{and}
\author[4]{Germ\'{a}n Lugones\href{https://orcid.org/0000-0002-2978-8079}{\includegraphics[scale=0.1]{ORCIDiD_icon128x128.png}}}

\affiliation[1]{Department of Physics, Tamkang University, New Taipei 251, Taiwan}

\affiliation[2]{Centro de Ci\^{e}ncias Exatas, Naturais e Tecnol\'{o}gicas, UEMASUL,  Rua Godofredo Viana 1300, Centro CEP: 65901-480, Imperatriz,  Maranh\~{a}o, Brazil}

\affiliation[3]{Department of Physics, Instituto Tecnologico de Aeronautica, DCTA , 12228-900, S\~{a}o Jos\'{e} dos Campos, SP, Brazil}

\affiliation[4]{Universidade Federal \rm{do} \rm{ABC}, Centro de Ci\^{e}ncias Naturais e Humanas, Avenida dos Estados, 5001,
 \rm{CEP} 09210-170, Santo Andr\'{e}, SP, Brazil}




\newcommand{\be}{\begin{equation}}
\newcommand{\ee}{\end{equation}}
\newcommand{\bea}{\begin{eqnarray}}
\newcommand{\eea}{\end{eqnarray}}
\newcommand{\red}{\textcolor{black}}

\abstract{We study quark-hadron hybrid stars with sharp phase transitions assuming that phase conversions at the interface are slow. Hadronic matter is described by a set of equations of state (EoS)  based on the chiral effective field theory and quark matter by a generic bag  model. Due to slow conversions at the interface, there is an extended region of stable hybrid stars with central densities above the density of the maximum mass star. We explore systematically the role of the transition pressure and the energy-density jump $\Delta \epsilon$ at the interface on some global properties of hybrid stars, such as the maximum mass,  the last stable configuration, and tidal deformabilities. We find that for a given transition pressure, the radius of the last stable hybrid star decreases as $\Delta \epsilon$ raises  resulting in a larger extended branch of stable hybrid stars.  Contrary to purely hadronic stars, the tidal deformability $\Lambda$ can be either a decreasing or an increasing function of the stellar mass $M$ and for large values of the transition pressure has a very weak dependence on $M$. Finally, we analyze the tidal deformabilities $\Lambda_1$ and $\Lambda_2$ for a binary system with the same chirp mass as GW170817.  In the scenario where at least one of the stars in the binary is hybrid, we find that models with low enough transition pressure are inside the  $90 \%$ credible region of GW170817. However, these models have maximum masses below $2 \, M_{\odot}$, in disagreement with observations. We also find that the LIGO/Virgo constrain (at $90\%$ level) and the  $2 \, M_{\odot}$ requirement can be simultaneously fulfilled in a scenario where all hybrid configurations have masses larger than  $1.6 \, M_{\odot}$ and the hadronic EoS is not too stiff, such as several of our hybrid models involving a hadronic EoS of intermediate stiffness. In such scenario hybrid stars may exist in Nature but both objects in GW170817 were hadronic stars. }


\maketitle

\flushbottom

\section{Introduction} \label{sec:intro}
The composition and properties of neutron stars (NSs), which are among the densest objects in the universe, are not completely understood theoretically.
However, measurements of the masses and radii of these objects can strongly constrain their equation of state (EoS) and consequently their internal composition (see  \cite{2012ARNPS..62..485L,2016PhR...621..127L} for a complete review). In fact,  the most stringent constraints for the EoS come from  the two solar mass limit of the pulsars PSR J1614-2230 \cite{2010Natur.467.1081D}, PSR J0348+0432 \cite{2013Sci...340..448A}, PSR J0740+6620  \cite{Cromartie:2019kug}, and PSR J2215-5135  \cite{Linares:2018ppq}. Mass and radius determinations are receiving a strong boost thanks to the observational data coming from the Neutron Star Interior Composition Explorer (NICER) experiment by NASA. NICER has provided a simultaneous measurement of both mass and radius of the millisecond-pulsar PSR J0030+0451 and also inferred some thermal properties of hot  regions present in the star \cite{2019ApJ...887L..21R,2019ApJ...887L..24M}.

The gravitational wave signal from mergers of binary NSs  \cite{2017PhRvL.119p1101A} is also sensitive to the EoS. New limits on the EoS were posed by the LIGO/Virgo detection of gravitational waves (GWs) originating from the NS-NS merger event GW170817 \cite{2017PhRvL.119p1101A},  which allowed determining the tidal deformabilities of the stars involved in the collision \cite{2018PhRvL.120q2703A,2018PhRvL.120q2702F,2018PhRvL.120z1103M,2018ApJ...857L..23R}.
During the inspiral phase of a NS-NS merger, strong tidal gravitational fields deform the multipolar structure of the NSs,
leaving a detectable imprint on the observed gravitational waveform of the merger \cite{2017PhRvL.119p1101A,2018PhRvD..98f3020Z}.
This effect can be quantified in terms of the so-called dimensionless tidal deformability of the star, defined as:
\be  \Lambda_i=\frac{2}{3}k_2^{(i)}\left(\frac{c^2 R_i}{GM_i}\right)^5 , \ee   
where $k_2^{(i)}$ is the second Love number, $R_i$ the radius, and $M_i$ the mass of the $i$-th star \cite{2008PhRvD..77b1502F}.  $\Lambda_i$ describes the amount of induced mass quadrupole moment when reacting to a certain external tidal field  (see
\cite{2008ApJ...677.1216H,2009PhRvD..80h4035D}  for more details).

Previously, several authors have studied tidal effects trying to impose some constraints on the EoS  \cite{2018PhRvC..98d5804T,2018PhRvC..98c5804M,2018PhRvL.121i1102D,2019PhRvD..99d3010C,2020arXiv200304471S}.
Some works analyzed the tidal properties of NSs using relativistic mean-field models \cite{2019PhRvC..99e2802N,2019PhRvC..99d5202L} and  Skyrme-type models \cite{2018PhRvC..98f5805K,2019PhLB..796....1T,2020EPJA...56...32L}. Other focused on the role of the symmetry energy $E_{\rm sym}(\rho)$ \cite{2019JPhG...46g4001K} and analyzed the role of the appearance of $\Delta$-isobars taking into account the data from GW170817 \cite{2019ApJ...874L..22L}.

Among  the open questions that may be explored with GW measurements, the possible appearance of quark matter in compact stars is an issue that has attracted the attention for decades. In fact, it has been conjectured long ago \cite{1965Ap......1..251I,1966Natur.209..389P,1970PThPh..44..291I} that a higher density class of compact stars may arise in the form of hybrid stars or quark stars (QSs), whose core or entire volume consists of quark matter (see also \cite{2014PhRvD..89j3014M,2015ApJ...815...81M,Lugones:2015bya} and references therein). Nowadays, it is generally believed that deconfined quarks may exist inside the core of compact stars, which corresponds to the high-density and low-temperature region of the QCD phase diagram.

An important issue related with the hadron--quark phase transition in compact star interiors, is the possible existence of \textit{mass twins} in the mass-radius diagram.  The term mass twins refers to the existence of doublets (and sometimes multiplets) of stellar configurations with the same gravitational mass but different radii. The appearance of twins may be a consequence of the occurrence of a  third family of  compact stars, besides white dwarfs and ordinary NSs. As already recognized long ago \cite{1968PhRv..172.1325G},  the third family is related to the behavior of the high-density EoS, which may exhibit a phase transition \cite{1983JPhG....9.1487K,Schertler:2000xq,2000A&A...353L...9G}. In this context,  EoSs with multiple phases which include a deconfined quark-matter segment were analysed as well \cite{2013PhRvD..88h3013A, 2015PhRvD..92h3002A, 2014JPhG...41l3001B, 2015A&A...577A..40B, 2016PhRvC..93d5812R, 2017PhRvD..96e6024K, 2018EPJA...54...28C} (see also  \cite{2018RPPh...81e6902B} for a recent review). Moreover, an additional phase transition in the quark core can lead to a fourth family of compact stars \cite{2017PhRvL.119p1104A}. The observation of twins or multiplets may be regarded as a signature of the existence of hybrid stars but other ways of identifying them would be possible.  For example,  general relativistic simulations of merging NSs including quarks at finite temperatures \cite{2019PhRvL.122f1101M}, show that the phase transition would lead to a post-merger signal considerably different from the one expected from the inspiral.

Besides the standard mass twins and multiplets described before, it has been shown recently that there is another mechanism that gives rise to an additional family of twins. In fact, in the case of hybrid stars having \textit{slow} quark-hadron conversion rates at a \textit{sharp} interface, there exists a new connected \textit{stable} branch beyond the maximum mass star that has $\partial M / \partial \epsilon_{\mathrm{c}}<0$, being $M$ the stellar mass and $\epsilon_{\mathrm{c}}$ the energy density at the stellar center \cite{Flores:2012vf, 2018ApJ...860...12P, 2019MNRAS.489.4261M}. In this paper we  analyze the possibility of identifying these new hybrid objects by means of the observation of their tidal deformabilities. 
%

This paper is organized as follows. In Sec.~\ref{sec:eos}, we describe the hadronic and quark matter EoSs used in this work and combine them in order to obtain hybrid EoSs with sharp first order phase transitions at some given pressures.
In Sec.~\ref{sec:slow}  we summarize the role of slow transitions on the dynamical stability of hybrid stars and in Sec.~\ref{sec:equations} we present the equations for the tidal deformabilty used in the calculations.
In Sec.~\ref{sec:tidal} we present our results for the tidal deformations of hybrid stars.
In Sec.~\ref{sec:conclusions} we present our main conclusions.

\section{Equations of State}\label{sec:eos}

In this section, we discuss the cold dense-matter models we use in our analysis. In Sec. \ref{sec:eos_H} we describe the hadronic model, in Sec. \ref{sec:eos_Q} the quark matter model, and in Sec.  \ref{sec:transitions} we explain the procedure used to construct first order phase transitions with a sharp density discontinuity.  

\subsection{Hadronic Matter}
\label{sec:eos_H}

For the hadronic phase (HP) we use the model presented in Ref. \cite{2013ApJ...773...11H},  which is based on nuclear interactions derived from chiral effective field theory (EFT). In recent years, the development of chiral EFT has provided the framework for a systematic expansion for nuclear forces at low momenta, allowing one to constrain the properties of neutron-rich matter up to nuclear saturation density to a high degree.
However, our knowledge of the EOS at densities greater than 1 to 2 times the nuclear saturation density is still insufficient due to limitations on both laboratory experiments and theoretical methods.
Because of this, the EoS at supranuclear densities is usually described by a set of three polytropes which are valid, respectively, in three consecutive density regions. For these polytropic EoSs it is required non-violation of causality and consistence with the recently observed pulsars with  $\sim 2 M_{\odot}$ \cite{2010Natur.467.1081D,2013Sci...340..448A}. In this work we will use three representative EoSs labeled as soft, intermediate and stiff that have been presented in Ref. \cite{2013ApJ...773...11H} . 
To describe the structure of the crust we follow the formalism developed in Ref. \cite{1971ApJ...170..299B}  for the outer crust, while we employ the pioneering work of Ref. \cite{1973NuPhA.207..298N} for the description of the inner crust.

\subsection{Quark Matter}
\label{sec:eos_Q}

For the quark phase (QP) we adopt a generic bag model defined by the following grand thermodynamic potential \cite{2005ApJ...629..969A}:
\be\label{quark1}   \Omega_{\rm QM}=-\frac{3}{4\pi^2}a_4\; \mu^4 +\frac{3}{4\pi^2}a_2 \;\mu^2 +B_{\rm eff}+\Omega_e,  \ee
where $\mu\equiv (\mu_u+\mu_d+\mu_s)/3$ is the quark chemical potential, $B_{\rm eff}$, $a_4$, and $a_2$ are  three free parameters independent of $\mu$,
and $\Omega_e$ is the grand thermodynamic potential for electrons $e$.
For hybrid systems, though, it happens that the contribution to the thermodynamic quantities coming from electrons is negligible
(see discussion in  \cite{2018ApJ...860...12P}).
The influence of strong interactions on the pressure of the free-quark Fermi sea is roughly taken into account by the parameter $a_4$, 
where $0\leq a_4\leq 1$, and $a_4=1$ indicates no correction to the ideal gas.
The effect of the color superconductivity phenomenon in the Color Flavor Locked (CFL) phase can be explored setting
$a_2=m_s^2-4\Delta^2$, being $m_s$ the mass of the strange quark and $\Delta$ the energy gap associated with quark pairing. 
The standard MIT bag model is obtained for $a_4=1$ and $a_2=m_s^2$. 
The bag constant $B_{\rm eff}$ is related to the confinement of quarks, representing in a phenomenological way the vacuum energy \cite{1984PhRvD..30.2379F}.
From Eq. (\ref{quark1}), one can obtain the pressure $p=-\Omega_{\rm QM}$, the baryon number density
\be\label{quark2}  n=-\frac{1}{3}\frac{\partial\Omega_{\rm QM}}{\partial \mu}=\frac{1}{2\pi^2}(2a_4 \mu^3-a_2\mu),  \ee
and the energy density
\bea\label{quark3}
 \epsilon &=& 3n\mu  - p = \frac{9}{4\pi^2}a_4\mu^4-\frac{3}{4\pi^2}a_2\mu^2+B_{\rm eff}\nonumber\\   
              &= & 3p+4B_{\rm eff}+\frac{3a_2}{2\pi^2}\mu^2. 
\eea
An analytic expression of the form $p=p(\epsilon)$ for this phenomenological model can be simply obtained if one solves Eq. (\ref{quark3}) for $\mu$
and replaces it in Eq. (\ref{quark1}). The final result is
\be\label{quark4}  p(\epsilon)=\! \frac{(\epsilon\!  -4B_{\rm eff})}{3}-\frac{a_2^2}{12\pi^2 a_4}\! \left[1\! +\! \sqrt{1+\frac{16\pi^2a_4}{a_2^2}(\epsilon-B_{\rm eff})}\right].  \ee

\subsection{Phase Transitions}
\label{sec:transitions}

\begin{table}[tbp]
\centering
\begin{tabular}{ccccccccc}
\hline\hline
Model &
 $p_{\rm{tr}}$ & 
$\epsilon_{\rm{tr}}^{\rm{HP}}$   & 
$n_{\rm B,tr}^{\rm{HP}}$  &  
$ \Delta\epsilon$ & 
$a_2^{1/2}$ & 
$B_{\rm eff}^{1/4}$ & 
$a_4$  &  
$M_{\rm{max}}$ \\
 &
$\left(\rm{\frac{MeV}{fm^{3}}}\right)$ & 
$\left(\rm{\frac{MeV}{fm^{3}}}\right)$ & 
$(\rm{fm^{-3}})$  &  
$\left(\rm{\frac{MeV}{fm^{3}}}\right)$ & 
$(\rm{MeV})$ & 
$(\rm{MeV})$ & 
 &  
$(M_\odot)$ \\
 \hline  \hline  
1   &  $77.9$     &  $585.2$   &   $0.592$  &  $60$      &  $100$    & $164.79$   &   $0.976$    &  $1.497$  \\
2   &  $124.3$   &  $640.7$   &   $0.639$  &  $50$      &  $100$    & $153.15$   &   $0.848$    &  $1.666$   \\                  
3   &  $124.3$   &  $640.7$   &   $0.639$  &  $120$    &  $100$    & $161.75$   &   $0.919$    &  $1.640$   \\
4   &  $124.3$   &  $640.7$   &   $0.639$  &  $200$    &  $100$    & $170.15$   &   $0.999$    &  $1.565$   \\
5   &  $165.5$   &  $719.1$   &   $0.704$  &  $100$    &  $100$    & $153.35$   &   $0.837$    &  $1.490$   \\
6   &  $165.5$   &  $719.1$   &   $0.704$  &  $250$    &  $100$    & $170.30$   &   $0.960$    &  $1.564$  \\
7   &  $242.8$   &  $844.6$   &   $0.799$  &  $300$    &  $100$    & $163.89$   &   $0.856$    &  $1.724$   \\
8   &  $242.8$   &  $844.6$   &   $0.799$  &  $400$    &  $100$    & $173.85$   &   $0.916$    &  $1.714$    \\                  
9   &  $242.8$   &  $844.6$   &   $0.799$  &  $500$    &  $100$    & $182.34$   &   $0.976$    &  $1.712$   \\
10 &  $359.8$   &  $1004$    &   $0.912$  &  $500$    &  $100$    & $163.87$   &   $0.783$    &  $1.866$   \\
11 &  $359.8$   &  $1004$    &   $0.912$  &  $700$    &  $100$    & $182.33$   &   $0.865$    &  $1.865$    \\
12 &  $359.8$   &  $1004$    &   $0.912$  &  $900$    &  $100$    & $196.45$   &   $0.947$    &  $1.865$    \\                    
13 &  $485.8$   &  $1154$    &   $1.008$  &  $600$    &  $100$    & $146.29$   &   $0.671$    &  $1.947$   \\
14 &  $485.8$   &  $1154$    &   $1.008$  &  $800$    &  $100$    & $170.35$   &   $0.730$    &  $1.946$    \\
15 &  $485.8$   &  $1154$    &   $1.008$  &  $1000$  &  $100$    & $187.13$   &   $0.788$    &  $1.946$    \\  
\hline\hline
\end{tabular}
\caption{We list our choices for the parameters $a_2$, $p_{tr}$ and $\Delta \epsilon$, which according to the discussion given in Sec. \ref{sec:transitions},  define the localisation of the quark-hadron interface as well as the parameters $B_{\rm eff}$    and  $a_4$ of the quark matter EoS. In this Table it is assumed that the hadronic EoS is described by the \textit{soft} model of Ref. \cite{2013ApJ...773...11H}. For completeness we include the maximum stellar mass $M_{\max }$ of each model.}
\end{table}

\begin{table}[tbp]
\centering
\begin{tabular}{ccccccccc}
\hline\hline
Model &
 $p_{\rm{tr}}$ & 
$\epsilon_{\rm{tr}}^{\rm{HP}}$   & 
$n_{\rm B,tr}^{\rm{HP}}$  &  
$ \Delta\epsilon$ & 
$a_2^{1/2}$ & 
$B_{\rm eff}^{1/4}$ & 
$a_4$  &  
$M_{\rm{max}}$ \\
 &
$\left(\rm{\frac{MeV}{fm^{3}}}\right)$ & 
$\left(\rm{\frac{MeV}{fm^{3}}}\right)$ & 
$(\rm{fm^{-3}})$  &  
$\left(\rm{\frac{MeV}{fm^{3}}}\right)$ & 
$(\rm{MeV})$ & 
$(\rm{MeV})$ & 
 &  
$(M_\odot)$ \\
 \hline  \hline  
1   &   $19.25$  &  $295.0$   &  $0.304$  & $100$  &  $100$  & $156.68$  &  $0.784$  &  $1.646$  \\
2   &   $19.25$  &  $295.0$   &  $0.304$  & $150$  &  $100$  & $162.58$  &  $0.874$  &  $1.541$  \\                  
3   &   $19.25$  &  $295.0$   &  $0.304$  & $200$  &  $100$  & $167.90$  &  $0.964$  &  $1.453$   \\
4   &   $49.02$  &  $380.9$   &  $0.384$  & $150$  &  $100$  & $161.74$  &  $0.791$  &  $1.622$  \\
5   &   $49.02$  &  $380.9$   &  $0.384$  & $220$  &  $100$  & $169.16$  &  $0.882$  &  $1.510$ \\
6   &   $49.02$  &  $380.9$   &  $0.384$  & $280$  &  $100$  & $174.82$  &  $0.961$  &  $1.428$  \\
7   &   $104.5$  &  $475.3$   &  $0.464$  & $200$  &  $100$  & $158.36$  &  $0.683$  &  $1.865$  \\
8   &   $104.5$  &  $475.3$   &  $0.464$  & $300$  &  $100$  & $169.27$  &  $0.767$  &  $1.831$  \\                  
9   &   $104.5$  &  $475.3$   &  $0.464$  & $400$  &  $100$  & $178.41$  &  $0.851$  &  $1.830$   \\
10 &   $224.5$  &  $649.7$   &  $0.592$  & $400$  &  $100$  & $158.46$  &  $0.569$  &  $2.270$  \\
11  &  $224.5$  &   $649.7$  &  $0.592$  & $550$  &  $100$  & $174.09$  &  $0.633$  &  $2.269$   \\    
12 &   $224.5$  &  $649.7$   &  $0.592$  & $700$  &  $100$  & $186.38$  &  $0.698$  &  $2.269$  \\                    
13 &   $377.6$  &  $827.9$   &  $0.704$  & $600$  & $100$   & $145.08$  &  $0.445$  &  $2.428$  \\
14 &   $377.6$  &  $827.9$   & $0.704$   & $800$  & $100$   & $169.59$  &  $0.492$  &  $2.428$  \\
15 &   $377.6$  &  $827.9$   & $0.704$   & $1000$& $100$   & $186.56$  &  $0.540$  &  $2.428$  \\ 
\hline\hline
\end{tabular}
\caption{Same as in Table 1 but assuming that the hadronic EoS is described by the \textit{intermediate} model of Ref. \cite{2013ApJ...773...11H}.}
\end{table}

\begin{table}[tbp]
\centering
\begin{tabular}{ccccccccc}
\hline\hline
Model &
 $p_{\rm{tr}}$ & 
$\epsilon_{\rm{tr}}^{\rm{HP}}$   & 
$n_{\rm B,tr}^{\rm{HP}}$  &  
$ \Delta\epsilon$ & 
$a_2^{1/2}$ & 
$B_{\rm eff}^{1/4}$ & 
$a_4$  &  
$M_{\rm{max}}$ \\
 &
$\left(\rm{\frac{MeV}{fm^{3}}}\right)$ & 
$\left(\rm{\frac{MeV}{fm^{3}}}\right)$ & 
$(\rm{fm^{-3}})$  &  
$\left(\rm{\frac{MeV}{fm^{3}}}\right)$ & 
$(\rm{MeV})$ & 
$(\rm{MeV})$ & 
 &  
$(M_\odot)$ \\
 \hline  \hline  
1   &   $20.39$   &  $249.2$   &   $0.256$   &  $80$    &  $100$  & $147.08$   &   $0.622$     &   $1.852$   \\
2   &   $20.39$   &  $249.2$   &   $0.256$   &  $130$  &  $100$  & $154.11$   &   $0.705$     &   $1.711$   \\                  
3   &   $20.39$   &  $249.2$   &   $0.256$   &  $180$  &  $100$  & $160.28$   &   $0.789$     &   $1.596$  \\
4   &   $52.49$   &  $302.2$   &   $0.304$   &  $100$  &  $100$  & $142.52$   &   $0.535$     &   $2.054$  \\
5   &   $52.49$   &  $302.2$   &   $0.304$   &  $200$  &  $100$  & $156.81$   &   $0.645$     &   $1.825$ \\
6   &   $52.49$   &  $302.2$   &   $0.304$   &  $300$  &  $100$  & $168.01$   &   $0.756$     &   $1.770$   \\
7   &   $92.63$   &  $361.4$   &   $0.352$   &  $250$  &  $100$  & $154.54$   &   $0.547$     &   $2.318$  \\
8   &   $92.63$   &  $361.4$   &   $0.352$   &  $300$  &  $100$  & $160.67$   &   $0.584$     &   $2.318$  \\                  
9   &   $92.63$   &  $361.4$   &   $0.352$   &  $350$  &  $100$  & $166.17$   &   $0.621$     &   $2.317$   \\
10 &   $152.9$   &  $448.9$   &   $0.416$   &  $300$  &  $100$  & $147.16$   &   $0.443$     &   $2.662$   \\
11 &   $152.9$   &  $448.9$   &   $0.416$   &  $400$  &  $100$  & $160.35$   &   $0.489$      &   $2.661$   \\
12 &   $152.9$   &  $448.9$   &   $0.416$   &  $500$  & $100$   & $170.91$   &   $0.536$     &   $2.661$    \\                    
13 &   $259.1$   &  $573.6$   &   $0.496$   &  $500$  & $100$   & $145.66$   &   $0.359$     &   $2.874$    \\
14 &   $259.1$   &  $573.6$   &   $0.496$   &  $600$  & $100$   & $159.19$   &   $0.385$     &   $2.874$    \\
15 &   $259.1$   &  $573.6$   &   $0.496$   &  $700$  & $100$   &  $169.95$   &  $0.411$     &   $2.874$   \\  
\hline\hline
\end{tabular}
\caption{Same as in Table 1  but assuming that the hadronic EoS is described by the \textit{stiff} model of Ref. \cite{2013ApJ...773...11H}.}
\end{table}


The density at which the hadron-quark phase transition occurs is not known, but it is expected to occur at some times the nuclear saturation density.
In a NS, such a transition can lead to two possible types of internal structure, depending on the surface tension between hadronic and quark matter 
\cite{1998PhRvC..58.2538I,2011PhRvC..83f8801E}.
If the surface tension between hadronic and quark matter is larger than a critical value, which is estimated to be around  $60 \,  \rm MeV/fm^2$ \cite{Voskresensky2003,Maruyama2007},  then there is a sharp interface and  we have a Maxwell phase transition.
If the surface tension is below the critical value, then there is a mixed phase of pure nuclear matter and pure quark matter usually called Gibbs phase transition.
In this work we assume that the interface between hadronic and quark matter is a sharp interface, which is a possible scenario given the uncertainties in the value of the surface tension \cite{2012PhRvC..86b5203P,2013PhRvC..88d5803L,Lugones:2016ytl,Lugones:2018qgu}.

At the quark-hadron interface the following conditions are verified:
\begin{itemize}
\item the pressure is continuous (mechanical equilibrium):
\be\label{transitions2}   p_{\rm{tr}}^{\rm{QP}}= p_{\rm{tr}}^{\rm{HP}} \equiv p_{\rm{tr}}.  \ee

\item The energy density has a jump $\Delta\epsilon$, i.e.: 
\be\label{transitions3}   \epsilon_{\rm{tr}}^{\rm{QP}}=\epsilon_{\rm{tr}}^{\rm{HP}}+\Delta\epsilon.  \ee

\item The Gibbs free energy per baryon at zero temperature, $g=(p +\epsilon)/n$, is continuous (chemical equilibrium):
\be\label{transitions4}    g_{\rm{tr}}^{\rm{QP}}=g_{\rm{tr}}^{\rm{HP}}.   \ee
\end{itemize}

From Eq. (\ref{quark3}) we see that $g^{\rm{QP}}=3\mu$, and for hadronic matter we write $g^{\rm{HP}}=(p^{\rm{HP}} +\epsilon^{\rm{HP}})/n^{\rm{HP}}$. Therefore, at the interface we have:
\be\label{transitions5}    3\mu_{\rm{tr}}=\frac{p_{\rm{tr}}^{\rm{HP}} + \epsilon_{\rm{tr}}^{\rm{HP}}}{n_{\rm{tr}}^{\rm{HP}}}.   \ee
Now using Eq. (\ref{quark1}), we can write Eq. (\ref{transitions2}) as:
\be\label{transitions6}    p_{\rm{tr}}^{\rm{HP}}=\frac{3}{4\pi^2}a_4 \mu_{\rm{tr}}^4 -\frac{3}{4\pi^2}a_2 \;\mu_{\rm{tr}}^2 -B_{\rm eff}.  \ee
Similarly, using Eq. (\ref{quark3}), we can write Eq. (\ref{transitions3}) as:
\be\label{transitions7}   \epsilon_{\rm{tr}}^{\rm{HP}}+\Delta\epsilon=3p_{\rm{tr}}^{\rm{HP}}+4B_{\rm eff}+\frac{3a_2}{2\pi^2}\mu_{\rm{tr}}^2,  \ee
where $\mu_{\rm{tr}}$ is given by Eq. (\ref{transitions5}).
From Eqs. (\ref{transitions6})--(\ref{transitions7}) we can write the parameters  $B_{\rm eff}$ and $a_4$ in terms of 
$p_{\rm{tr}}^{\rm{HP}}$, $ \epsilon_{\rm{tr}}^{\rm{HP}}$, $n_{\rm tr}^{\rm{HP}}$, $\mu_{\rm{tr}}$, $\Delta\epsilon$, and $a_2$  in the form:
\be
\left\{ 
\begin{array}{rl}\label{system}
 B_{\rm eff} &= \beta_0+ \beta_1\Delta\epsilon +  \beta_2 a_2\\
a_4 & = \alpha_0+  \alpha_1\Delta\epsilon +  \alpha_2 a_2
\end{array}
\right.
\ee
with
$$ \; \; \; \beta_0=\frac{1}{4}( \epsilon_{\rm{tr}}^{\rm{HP}}-3p_{\rm{tr}}^{\rm{HP}}), \; \; \; \; \; \;  \;     \beta_1=\frac{1}{4},  \; \; \; \; \;  \;  \; \;   \beta_2=-\frac{3\mu_{\rm{tr}}^2}{8\pi^2}, $$
$$\alpha_0=\frac{\pi^2}{3\mu_{\rm{tr}}^4}( \epsilon_{\rm{tr}}^{\rm{HP}}+p_{\rm{tr}}^{\rm{HP}}), \; \; \; \; \;     \alpha_1=\frac{\pi^2}{3\mu_{\rm{tr}}^4},   \; \; \; \;  \alpha_2=\frac{1}{2 \mu_{\rm{tr}}^2}.  $$
These coefficients are fixed once the transition point has been chosen.

\begin{figure}[ht!]
\centering
\includegraphics[width=7.8cm]{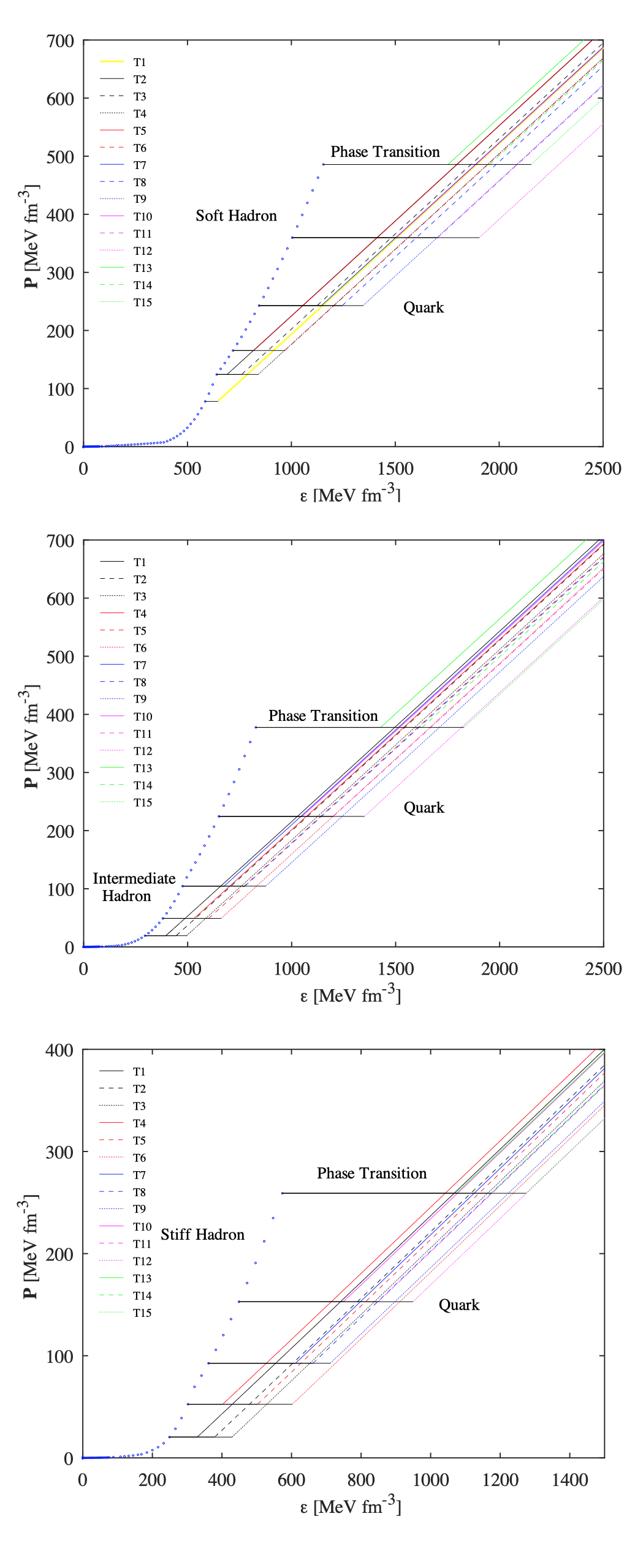}   
\caption{Hybrid EoSs used in this work. For each of the three hadronic EoS labeled as soft, intermediate and stiff we consider fifteen possible phase transitions using different values of the transition pressure $p_{tr}$ and the density jump $\Delta \epsilon$ at the interface (see  Sec.~\ref{sec:transitions}). The numerical values of the parameters for each model can be found in Tables 1, 2 and 3.}
\label{fig:eos1}
\end{figure}

In this work we fix the parameter  $a_2$ to the typical value  $a_2^{1/2}= 100 \, \rm{MeV}$ and explore several values of the transition pressure $p_{\rm{tr}}$  and the density jump $\Delta\epsilon$ at the interface,  spanning a wide range of values for these quantities. Our choices for $p_{\rm{tr}}$  and $\Delta\epsilon$ are shown in Tables 1, 2 and 3, together with other properties of matter at the discontinuity, the corresponding values of the quark matter EoS parameters, and the maximum mass $M_{\max }$ for each model.  The values presented in Tables 1, 2 and 3  correspond, respectively, to the soft, intermediate and stiff hadronic EoSs of Ref. \cite{2013ApJ...773...11H}.  The curves of the resulting hybrid EoSs are shown in Fig. \ref{fig:eos1}.

\section{Slow transitions and dynamical stability of hybrid stars}
\label{sec:slow}

When matter in the neighborhood of the quark-hadron interface is perturbed and oscillates around an equilibrium position, reactions between both phases may occur due to compression and rarefaction of fluid elements. The physics of these reactions can be classified in slow or fast depending on their timescale.  Slow transitions are those with a timescale  greater than the timescale of the oscillations. In the opposite case, when the timescale of the reactions is lower than the timescale of the oscillations, we have a fast transition. In the slow case, when fluid elements near the interface are perturbed and displaced from their equilibrium position, they  maintain their composition and co-move with the motion of the interface.
In this work we will focus our discussion on the consequences of slow transitions.
A slow phase transition has significant consequences at the macroscopic level, when the dynamical behavior of the star against perturbations is investigated \cite{2018ApJ...860...12P,2019MNRAS.489.4261M}. The effect can be encoded in the junction conditions for the radial eigenfunctions $\xi$ and $\Delta p$. In particular, in the case of slow transitions they are continuous variables, which can be written as
\begin{equation}
    [\xi]= \xi_{+}-\xi_{-}=0,\quad \quad [\Delta p] = (\Delta p)_{+}-(\Delta p)_{-}=0
\end{equation}
where $[x] \equiv x_{+}-x_{-}$, with $x_{+}$ and $x_{-}$ defined as the values after and before the interface, taking a reference system centered at the star. One direct consequence  of the discontinuity conditions for the slow transition case is that the region of stable  stars can be extended beyond the  maximum mass point. This can be shown by determining the frequencies of the fundamental radial oscillation modes \cite{Flores:2012vf,2018ApJ...860...12P,2019MNRAS.489.4261M}. 
In general,  the frequency $\omega_{0}$ of the fundamental mode verifies $(\omega_{0}^2 > 0)$ for the dynamically stable configurations, and the last stable object has $\omega_{0}=0$. Coincidentally, for homogeneous (one phase) stars, the last stable star is the one that is just localized at the top of the mass versus radius curve (the maximum mass point). This fact justifies the wide use of the following static stability criterion instead of the dynamical one:
\begin{equation}
    \frac{\partial M}{\partial \epsilon_c}<0 \quad \Rightarrow \quad   \text{(unstable configuration)},
\end{equation}
\begin{equation}
    \frac{\partial M}{\partial \epsilon_c}>0 \quad  \Leftarrow  \quad  \text{(stable configuration)}.
\end{equation}
In practice, this criterion is used to identify the last stable stellar configuration because, for a one-phase star, when $\omega_{0}=0$ the derivative $\partial M / \partial \epsilon_c$ changes its sign.

\begin{figure}[ht!]
\centering
\includegraphics[width=7.8cm]{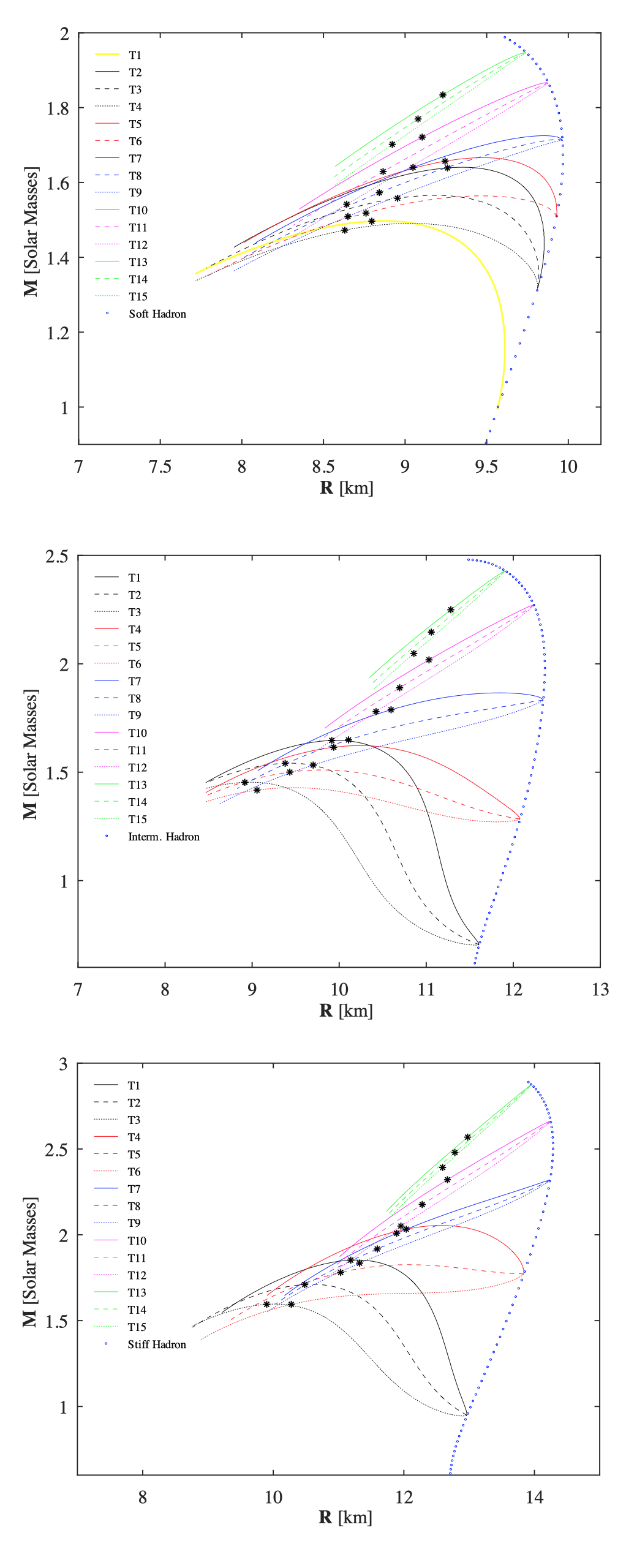} 
\caption{Mass-radius relations for the hybrid EoSs shown in Fig.~\ref{fig:eos1}.  The asterisks represent the last stable star of each model, having a zero frequency fundamental radial mode. As explained in Sec.~\ref{sec:slow}, since we assume  that quark-hadron phase conversions at the interface of a hybrid star have a slow timescale, there are parts of the curves that are dynamically stable in spite of having $\partial M / \partial \epsilon_c < 0$.} 
\label{fig:massvsr}
\end{figure}

However, in the case of two-phase stars with slow conversions at the interface, zero frequency modes happen in general after the point of maximum mass (for further details see e.g. Ref. \cite{2018ApJ...860...12P}). This gives rise to an extended region of stable hybrid stars with central densities that are larger than the density of the maximum mass star, as can be seen in Fig. \ref{fig:massvsr}. The last stable star, which has a zero frequency fundamental mode, has been identified in Fig. \ref{fig:massvsr} with an asterisk. Notice that hybrid models with low enough transition pressure are not able to reach a maximum mass above the observed value $M_{\max} \approx  2 M_{\odot}$. As apparent from Fig. \ref{fig:massvsr}, for a given transition pressure, the largest the density jump $\Delta \epsilon$, the smallest the radius of the last stable hybrid star.

\section{Tidal Deformability equations}
\label{sec:equations}

The theory of relativistic tidal effects in binary systems has been focus of intense research in recent years \cite{2009PhRvD..80h4035D, poisson2009, 2008PhRvD..77b1502F, Fattoyev2013, 2018PhRvC..98c5804M, Hornick2018, Kumar2017, Hinderer2010, 2008ApJ...677.1216H}. In this section we will summarize our procedure for computing the important physical quantities.

The first step in order to investigate the tidal deformability of binary systems is the computation of the compact star structure by means of the  Tolman-Oppenheimer-Volkoff (TOV) equations, because some equilibrium quantities are needed during the numerical solution of the differential equations corresponding to the physics of tidal deformations. 

The tidal Love number $k_2$ is found using the following expression \cite{2008ApJ...677.1216H}:
\bea
k_2\!  &=& \!  \frac{8\mathcal{C}^5}{5}(1-2\mathcal{C})^2[2-y_R+2\mathcal{C}(y_R-1)] \nonumber\\ 
            &  &   \times \{ 2 \mathcal{C} [6-3y_R+3\mathcal{C}(5y_R-8)] \nonumber\\ 
            &  &  +4\mathcal{C}^3[13-11y_R +\mathcal{C}(3y_R-2)+2\mathcal{C}^2(1+y_R)] \nonumber\\ 
            &  &   +3(1\! -\! 2\mathcal{C})^2[2\! -\! y_R\!+\! 2\mathcal{C}(y_R\! -\! 1)]\! \log(1\! -\! 2\mathcal{C})\!  \}^{-1}
\eea
where $\mathcal{C}\equiv M/R$ is the dimensionless compactness parameter and $y_R\equiv y(R)$, being $y(r)$ the solution of the following first-order differential equation (see Ref. \cite{2010PhRvD..82b4016P} for more details):   
\begin{equation}
r \frac{dy(r)}{dr}+y(r)^2+y(r)F(r)+r^2Q(r)=0 .
\label{tidal}
\end{equation}
In this equation, the coefficients are: 
\be F(r)=[1-4\pi r^2(\epsilon-p)]\left[1-\frac{2m}{r}\right]^{-1}, \ee
\bea\label{functionQ}
Q(r) &=& 4\pi\left[5\epsilon +9p +\frac{\epsilon+p}{c_s^2}-\frac{6}{4\pi r^2}\right]\left[1-\frac{2m}{r}\right]^{-1} \nonumber\\ 
& &  -\frac{4m^2}{r^4}\left[1+\frac{4\pi r^3 p}{m}\right]^2\left[1-\frac{2m}{r}\right]^{-2}, 
\eea
where $c_s^2 \equiv dp/d\epsilon $ is the squared speed of sound. The boundary condition for Eq. (\ref{tidal}) at $r=0$ is given by  $y(0)=2$. In summary, the tidal Love number can be obtained once an EoS is supplied and the TOV equations together with Eq. (\ref{tidal}) are integrated.

\begin{figure}[ht!]
\centering
\includegraphics[width=8cm]{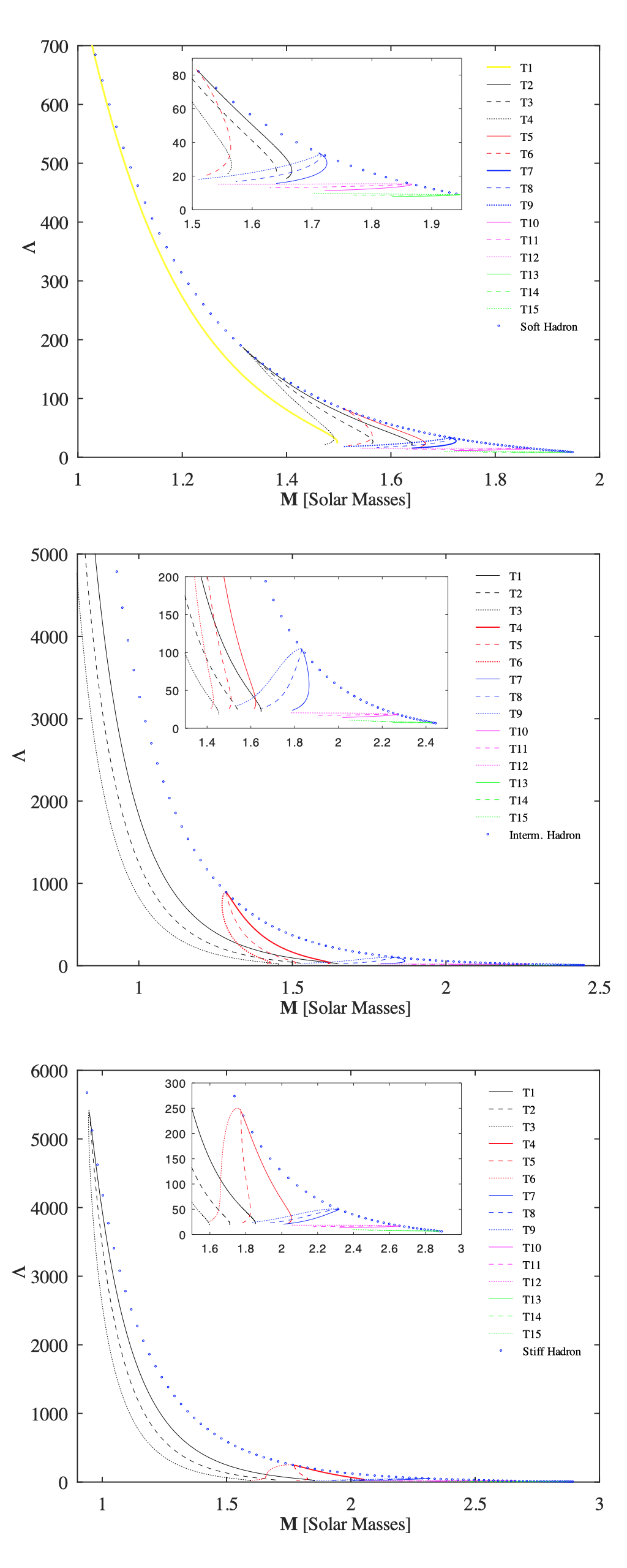} 
\caption{Tidal deformability $\Lambda$ as a function of the stellar mass for the hybrid EoSs presented in Fig.~\ref{fig:eos1}. 
\label{fig:tidal}}
\end{figure}

\section{Impact of phase transitions on tidal polarizability}
\label{sec:tidal}

Several models of NSs with first-order phase transitions have been recently analyzed (see e.g. Refs. \cite{2018PhRvD..97h4038P, 2018ApJ...857...12N, 2018ApJ...860..139B, 2019PhRvD..99b3009C, 2019A&A...622A.174S, 2019PhRvD..99h3014H, 2019PhRvD..99j3009M, 
 2019ApJ...877..139G, 2019MNRAS.489.4261M, 2020PhRvD.101d4019C}).
In the presence of a finite energy density discontinuity, the last term in Eq. (\ref{functionQ}) involves a singularity $\propto(\epsilon+p)/c_s^2$. This was first discussed by addressing the surface vacuum discontinuity for incompressible stars \cite{2009PhRvD..80h4035D}.  Then,  this approach was extended to the context of possible first-order transitions inside the stars  and  a junction condition was derived  \cite{2010PhRvD..82b4016P} that was corrected more recently \cite{Takatsy:2020bnx,Zhang:2020pfh}. In summary, the following  junction condition for the $y(r)$ function has to be imposed:
\begin{equation}
y(r_{d} +\epsilon)=y(r_{d} -\epsilon)   -\frac{\Delta \rho}{\tilde{\rho} / 3 + p\left(r_{d}\right)}
\label{eq:junction_new}
\end{equation}
where $r_{d}$ represents the coordinate radius of the interface, i.e. the point where the phase transition takes place. The region $r< r_{d}$ represents the internal core containing quark matter and $r>r_{d}$ is the external hadronic region of the star. We also have that $\tilde{\rho}= m(r_{d})/(4\pi r_{d}^{3}/3)$ and $\Delta \rho = \rho(r_{d}+\epsilon) - \rho(r_{d}-\epsilon)$ \footnote{Eq. (\ref{eq:junction_new}) contains an extra $p(r_{d})$ term in the denominator as compared to Eq. (14) of Ref. \cite{2010PhRvD..82b4016P} as shown in Refs. \cite{Takatsy:2020bnx,Zhang:2020pfh}. We have checked that in our case  the original and the corrected formulae lead to differences in the tidal deformabilities that are less that $1.5 \%$.}.

The accumulated phase contribution due to the deformation from both of the stars is included in the inspiral  signal as the combined dimensionless tidal deformability, which is given by
\be  \tilde{\Lambda}=\frac{16}{13} \left[\frac{(m_1+12m_2)m_1^4\Lambda_1+(m_2+12m_1)m_2^4\Lambda_2}{(m_1+m_2)^5}\right],\ee
where $\Lambda_1(m_1)$ and $\Lambda_2(m_2)$ are the tidal deformabilities of the individual binary components \cite{2014PhRvL.112j1101F}.
The quantity $\tilde{\Lambda}$ is usually evaluated as a function of the chirp mass $\mathcal{M}_c=(m_1 m_2)^{3/5}/(m_1+m_2)^{1/5}$, for various values of the mass ratio $q=m_2/m_1$. 
From the event GW170817, it was possible to set the first upper limit on the dimensionless tidal deformability  $\Lambda_{1.4}$ of a NS with a mass $1.4M_{\odot}$ such that $\Lambda_{1.4}\leq 800$ at $90\%$ confidence level (for the case of low-spin priors) \cite{2017PhRvL.119p1101A}. Independently of the waveform model or the choice of priors, the source-frame chirp mass was constrained to the range $\mathcal{M}=1.188_{-0.002}^{+0.004} M_{\odot}$ and the mass ratio $q=m_2/m_1$ for low spin priors to the interval between $0.7-1.0$ \cite{2017PhRvL.119p1101A}.
Working under the hypothesis that both NSs are described by the same EoS and have spins within the range observed in Galactic binary NSs, the tidal deformability $\Lambda_{1.4}$ of a  $1.4 {M}_{\odot}$ NS was estimated through a linear expansion of $\Lambda(m) m^{5}$ around $1.4 {M}_{\odot}$ to be in the range $70-580$ at the $90 \%$ level \cite{2018PhRvL.121p1101A}.

\begin{figure}[ht!]
\centering
\includegraphics[width=8cm]{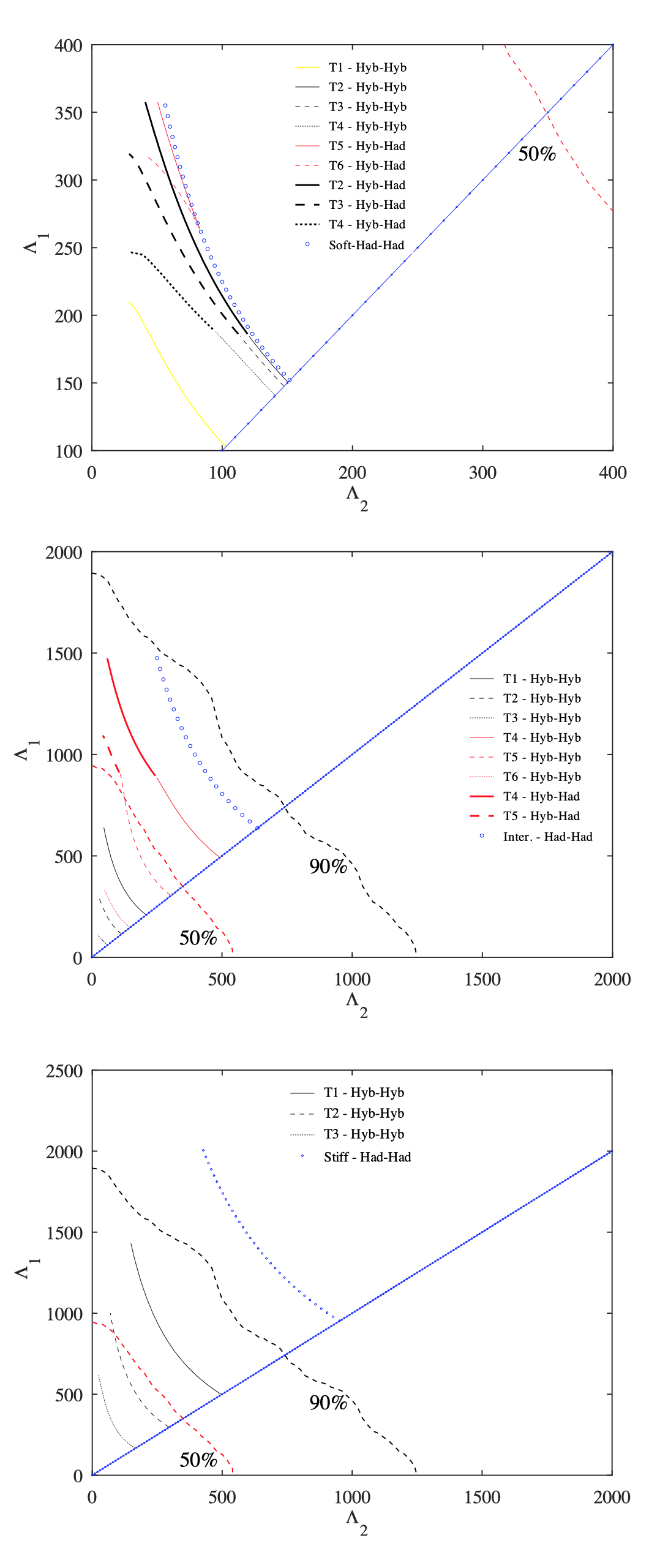} 
\caption{Dimensionless tidal deformabilities $\Lambda_1$ and $\Lambda_2$ for a binary NS system with masses $m_1$ and  $m_2$ and the same chirp mass as GW170817 \cite{2017PhRvL.119p1101A}. By definition we only plot combinations with $m_{1}>m_{2}$. The diagonal blue line indicates the $\Lambda_1 = \Lambda_2$ boundary.  The black dashed line denotes the $90\%$ credibility level and the red dashed line the $50\%$ level determined by LIGO/Virgo in the low-spin prior scenario.
\label{fig:tidaltidal}}
\end{figure}

In Fig. \ref{fig:tidal}, we show the dimensionless tidal deformability $\Lambda$ as a function of the stellar mass for all the hybrid models presented in Fig. \ref{fig:eos1}. Notice that $\Lambda$ is a decreasing function of $M$ for the hadronic branch. However, for hybrid configurations, $\Lambda$ can be either a decreasing or an increasing function of $M$. Moreover, for large values of the transition pressure, $\Lambda$ shows a very weak dependence on the stellar mass (the curves are quite horizontal).

In Fig. \ref{fig:tidaltidal}, we show the dimensionless tidal deformabilities $\Lambda_1$ and $\Lambda_2$ for a binary NS system with the same chirp mass as GW170817. The curves are obtained as follows: once a value of $m_1$ is chosen  one  can calculate $m_2$ by fixing $\mathcal{M} = 1.188\, M_{\odot}$. Running $m_1$ in the range of $1.36 \leq m_1 / M_{\odot} \leq 1.60$ we obtain  $m_2$ (with $1.17  \leq m_2 / M_{\odot} \leq 1.36$).  The values of $\Lambda_1$ and $\Lambda_2$ associated with $m_1$ and $m_2$ are depicted in Fig. \ref{fig:tidaltidal}.

We have analysed the three possibilities for the internal composition of the two stars in the binary system: hadron-hadron, hybrid-hadron and hybrid-hybrid. In the case with two purely hadronic stars, the presence of these binaries in the detection region depends strongly on the stiffness of the EoS. As can be seen in Fig. \ref{fig:tidaltidal} the soft hadronic EOS is inside the $50 \%$ region of GW170817, the intermediate one is inside the $90 \%$ region and the stiff one is outside the $90 \%$. Note that the hybrid models that we do not show in Figure \ref{fig:tidaltidal} are the cases that do not present hybrid configurations in the range 1.365 - 1.6 $M_{\odot}$ (these cases are completely represented in the $\Lambda_1$--$\Lambda_2$ plane by the hadronic curve with circle dots).  In the scenario where at least one of the stars in the binary is hybrid, we find that only models T1, T2, T3, T4, T5 and T6 fall inside the detection region. Some of these curves have a lower part corresponding to hybrid-hybrid mergers and an upper part corresponding to hybrid-hadron mergers. There are also curves where the lower part represents hadron-hadron mergers and the upper part hybrid-hadron mergers. In general, we observe that as the density jump $\Delta \epsilon$ increases, the curves shift to lower values of $\Lambda$. 

Notice that binary systems with at least one hybrid star that fall inside the  area of the $90 \%$ credible region for $\Lambda_1$--$\Lambda_2$ (T1, T2, T3, T4, T5 and T6) have maximum masses below $2 \, M_{\odot}$, and are therefore inconsistent with stringent constrains on the maximum NS mass coming from PSR J0348+0432 with $2.01 \pm 0.04 \, M_{\odot}$ \cite{2013Sci...340..448A}, PSR J0740+6620 with $2.14_{-0.09}^{+0.10} M_{\odot}$  \cite{Cromartie:2019kug}, and PSR J2215-5135 with  $2.27_{-0.15}^{+0.17} M_{\odot}$ \cite{Linares:2018ppq}. Therefore, for our model to agree with both the $90\%$ level determined by LIGO/Virgo and the $2 \, M_{\odot}$ pulsars, both objects in GW170817 must be hadronic stars and hybrid configurations cannot fall in the range 1.365 - 1.6 $M_{\odot}$. Viable candidates for this scenario would be the models from T7 to T15 linked to the hadronic EoS of \textit{intermediate} stiffness. These models have $M_{\max} > 2 \, M_{\odot}$ (see central panel of Fig. \ref{fig:massvsr}) and are represented in the $\Lambda_1$--$\Lambda_2$ plane by the hadronic curve with circle dots shown in the central panel of Fig. \ref{fig:tidaltidal}.

\section{Summary and conclusions}
\label{sec:conclusions}

In this work we have analyzed hybrid stars containing sharp phase transitions between hadronic and quark matter assuming that phase conversions at the interface are slow. Hadronic matter has been described by an EoS based on nuclear interactions derived from chiral effective field theory \cite{2013ApJ...773...11H} and quark matter by a generic bag  model \cite{2005ApJ...629..969A}. 

In the case of slow conversions at the quark-hadron interface, zero frequency radial modes arise in general beyond the point of maximum mass, giving rise to an extended region of stable hybrid stars with central densities that are larger than the density of the maximum mass star \cite{2018ApJ...860...12P}. This effect can be seen clearly in Fig. \ref{fig:massvsr} where large extended branches arise. We have explored systematically the role of the transition pressure $p_{\rm tr}$ and the energy-density jump $\Delta\epsilon$ on the maximum mass object, the last stable configuration, and the tidal deformabilities.
As expected, hybrid models with low enough $p_{\rm trans}$ are not able to reach a maximum mass above the observed value $M_{\max} \approx  2 M_{\odot}$. Also, for a given transition pressure, the radius of the last stable hybrid star decreases as $\Delta \epsilon$ raises  resulting in a larger extended branch of stable hybrid stars (see Fig. \ref{fig:massvsr}). 

We also found that for hybrid configurations, the tidal deformability $\Lambda$ can be either a decreasing or an increasing function of $M$.  Moreover, for large values of the transition pressure, $\Lambda$ shows a very weak dependence on the stellar mass and the curves become almost horizontal in some cases (see Fig. \ref{fig:tidal}). This is in contrast with the case of purely hadronic stars for which $\Lambda$ is always a decreasing function of $M$.

Finally, we analyzed the tidal deformabilities $\Lambda_1$ and $\Lambda_2$ for a binary NS system with the same chirp mass as GW170817 (see Fig. \ref{fig:tidaltidal}).  In the scenario where at least one of the stars in the binary is hybrid, we found that only models with low enough transition pressure fall inside the  $90 \%$ credible region for $\Lambda_1$--$\Lambda_2$.  Unfortunately, these models have maximum masses below $2 \, M_{\odot}$, i.e. in disagreement with the observation of  PSR J0348+0432, PSR J0740+6620  and PSR J2215-5135.  However, our model can explain the  $90\%$ level determined by LIGO/Virgo together with the $2 \, M_{\odot}$ constrain if both objects in GW170817 are hadronic stars and all hybrid configurations have masses larger than   $1.6 \, M_{\odot}$. We find that several hybrid models involving the hadronic EoS of intermediate stiffness are in agreement with observations.

\acknowledgments
Alessandro Parisi is grateful to Professor Feng-Li Lin  for many helpful discussions. 
Germ\'{a}n Lugones acknowledges the Brazilian agency Conselho Nacional de Desenvolvimento Cient\'{\i}fico e Tecnol\'{o}gico (CNPq) for financial support.
César H. Lenzi is thankful to the Fundação de Amparo à Pesquisa do Estado de São Paulo (FAPESP) under thematic project 2017/05660-0 for financial support. 
Chian-Shu Chen is supported by Ministry of Science and Technology (MOST), Taiwan, R.O.C. under Grant No. 107-2112-M-032-001-MY3.

\bibliographystyle{JHEP}
\bibliography{draft.bib}

\end{document}